\documentclass[twocolumn,prl]{revtex4}

\usepackage[dvips]{graphicx}
\begin{document}

\title{Universal charge transport of the Mn oxides in the high temperature limit}

\author{
W. Kobayashi$^1$, I. Terasaki$^{1,2}$, M. Mikami$^2$,
R. Funahashi$^{2,3}$, T. Nomura$^4$, and T. Katsufuji$^4$
}

\affiliation{
$^1$Department of Applied Physics, Waseda University, 
Tokyo 169-8555, Japan
}

\affiliation{
$^2$ CREST, Japan Science and Technology Agency, Tokyo 108-0075, Japan
}

\affiliation{
$^3$National Institute of Advanced Industrial Science and Technology, 
Osaka 563-8577, Japan
}

\affiliation{
$^4$Department of Physics, Waseda University, 
Tokyo 169-8555, Japan
}

\date{\today}

\begin{abstract}
We have found that various Mn oxides have the universal resistivity and thermopower 
in the high temperature limit. 
The resistivities and thermopowers of all the samples go 
toward constant values of
7$\pm$1 m$\Omega$cm and $-$79$\pm$3 $\mu$V/K, 
which are independent of carrier density and crystal structures.
We propose that the electric conduction occurs in a highly 
localized way in the high temperature limit, 
where the exchange of entropy and charge occurs 
in the neighboring Mn$^{3+}$ and Mn$^{4+}$ ions.
\end{abstract}

\maketitle

\section{Introduction}
In condensed matter physics, many researches 
have focused on low-temperature physics. 
Superconductivity, Anderson localization, superfluidity, 
metal-insulator transition, 
and Kondo effect are typical low-temperature physics. 
In contrast, studies on high-temperature physics are fewer 
in number at present. 

Recently, magnetic oxides have been studied extensively 
as a possible candidate for thermoelectric materials at high temperatures
since the discovery of large thermopower in NaCo$_2$O$_4$ 
\cite{terasaki, review}. 
The high thermoelectric performance of NaCo$_2$O$_4$ is currently
explained in terms of a large degeneracy of 
spin and orbital degrees of freedom of Co$^{4+}$ ions \cite{koshibae}. 
Therefore, control of spin and orbital states of magnetic ions 
would improve the performance of thermoelectric oxides. 
In this sense, thermoelectrics can be a new research direction
in magnetics. 

We hope that new physical phenomena such as the good thermoelectricity
in NaCo$_2$O$_4$ might exist in magnetic materials at high temperatures. 
Here we have paid attention to the transport properties of Mn oxides,
where a complicated electron-spin-phonon interaction
induces an exotic electric conduction such as colossal magnetoresistance.
We have measured their resistivity and thermopower up to 1000 K,
and have found that they are essentially independent 
of carrier density and crystal structure 
in the high temperature limit. 
This is quite anomalous in the sense that 
thermopower and resistivity usually depend on 
carrier density and crystal structure
for conventional metals and semiconductors.

\section{Experiment}
Polycrystalline samples of Ca(Mn$_{3-x}$Cu$_x$)Mn$_4$O$_{12}$ 
($x=$ 0, 0.25, 0.5, 0.75 and 1) and LiMn$_2$O$_4$ were prepared 
by a solid-state reaction. 
Stoichiometric amounts of CaCO$_3$, Mn$_3$O$_4$, CuO, and Li$_2$CO$_3$ 
were mixed, and the mixture was calcined 
at 900$^{\circ}$C for 12 h for CaMn$_3$Mn$_4$O$_{12}$, 
870$^{\circ}$C for 12 h for Ca(Mn$_{3-x}$Cu$_x$)Mn$_4$O$_{12}$ 
($x>$0), and 800$^{\circ}$C for 24 h for LiMn$_2$O$_4$. 
The product was finely ground, pressed into a pellet, 
and sintered at 950$^{\circ}$C for 72 h for CaMn$_3$Mn$_4$O$_{12}$, 
930$^{\circ}$C for 12 h for Ca(Mn$_{3-x}$Cu$_x$)Mn$_4$O$_{12}$ 
($x>$0), and 850$^{\circ}$C for 24 h for LiMn$_2$O$_4$. 
All the samples were calcined in air. 

Single crystals of Pr$_{0.5}$Ca$_{0.5}$MnO$_3$ were grown 
using the traveling solvent floating zone method. 
Stoichiometric amounts of Pr$_6$O$_{11}$, CaCO$_3$, and 
MnO$_2$ were mixed, and the mixture was calcined at 1200$^{\circ}$C for 8h 
in air. 
The calcined powder was ground, 
placed into a rubber tube and then hydrostatically 
pressed under 2000 kg/cm$^2$. 
The pressed rod was sintered at 1350$^{\circ}$C for 12 h in air. 
The feed and seed rods were rotated in opposite directions 
at a relative rate of 40 rpm and the melted zone was vertically 
scanned at a speed of 7 mm/h. 
The growth atmosphere was air. 

The X-ray diffraction (XRD) of the samples was measured using 
a standard diffractometer with Fe K$_\alpha$ radiation as an X-ray source 
in the $\theta$-$2\theta$ scan mode. 
The resistivity was measured by a four-probe method, 
below room temperature (4.2-300 K) in a liquid He cryostat, 
and above room temperature (300-1000 K) in an electric furnace. 
The thermopower was measured using a steady-state technique, 
below room temperature (4.2-300 K) in a liquid He cryostat, 
and above room temperature (300-1080 K) in an electric furnace. 
A temperature gradient of 0.5 K/mm was generated 
by a small resistive heater pasted on one edge of a sample below 500 K and 
by cooling one edge of a sample by an air pump above 500 K, 
and was monitored by a differential thermocouple made 
of copper-constantan below room temperature, 
and by that of platinum-rhodium above room temperature. 
The thermopower of the voltage leads was carefully subtracted. 

\begin{figure}[t]
\includegraphics[width=.8\linewidth]{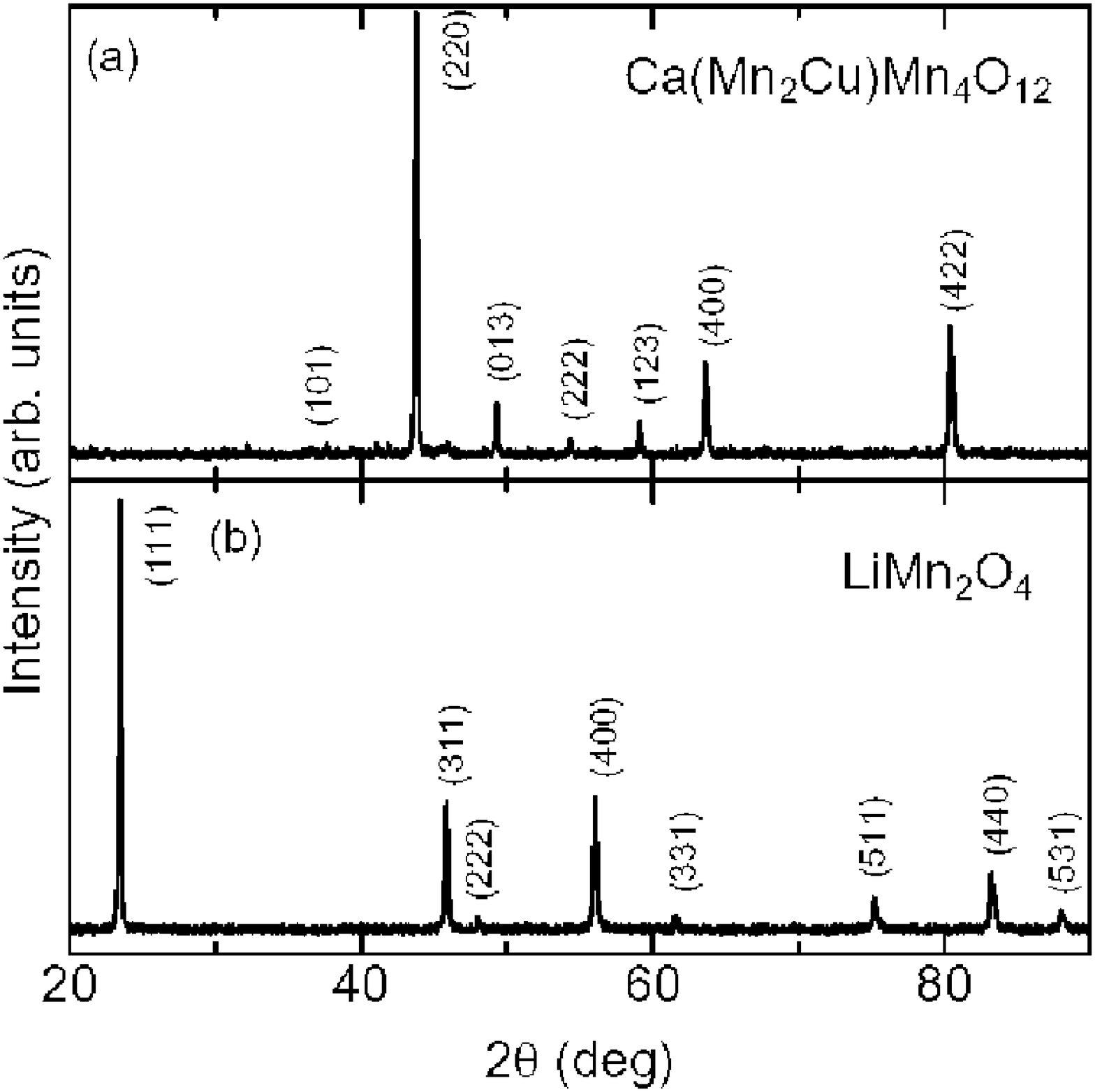}
\caption{
X-ray diffraction patters of
(a) Ca(Mn$_2$Cu)Mn$_4$O$_{12}$ and (b) LiMn$_2$O$_4$.
}
\end{figure}

\section{Results and Discussion}
Figure 1 shows typical X-ray diffraction patterns of 
(a) Ca(Mn$_2$Cu)Mn$_4$O$_{12}$ and (b) LiMn$_2$O$_4$. 
The samples were successfully prepared as single phases, 
and all the peaks are indexed \cite{bochu, LiMn}. 
Although crystal structures are different among
Ca(Mn$_{3-x}$Cu$_x$)Mn$_4$O$_{12}$ (CaCu$_3$Ti$_4$O$_{12}$-type), 
Pr$_{0.5}$Ca$_{0.5}$MnO$_3$ (GdFeO$_3$-type)
and LiMn$_2$O$_4$ (spinel),
they share a common feature that the Mn ions in the conduction path are 
either trivalent (Mn$^{3+}$) or tetravalent (Mn$^{4+}$) in the 
high-spin state, being surrounded by the oxygen octahedron.
The Mn$^{3+}$ ion has one $d$ electron in the doubly degenerate 
$e_g$ level, and the oxygen octahedron is elongated in one direction 
to release the degeneracy through the Jahn-Teller effect.
In contrast, the Mn$^{4+}$ ion has no Jahn-Teller effect, 
and is stable in the undistorted octahedron.
Thus, when a hole hops from Mn$^{4+}$ to Mn$^{3+}$, 
it inevitably accompanies the elongation of the oxygen octahedra,
forming a small polaron.

Figure 2 shows the resistivities of the Mn oxides. 
An important feature of the small-polaron conductor is that 
the mobility obeys an activation-type temperature dependence,
where the activation energy is the binding energy of the polaron
(the Jahn-Teller distortion energy or the $e_g$ level splitting
in the present case).
Then the resistivity is described by the activation transport as
\begin{equation}
 \rho=\rho_0\exp(E_{\rho}/k_BT),
\end{equation}
where $E_{\rho}$ is the activation energy seen in the resistivity.
As is clearly shown in Fig. 2, all the resistivities are well described 
by the activation process.
Some of the resistivities show an abrupt increase below a certain temperature,
which is due to the charge-ordering transition.
The transition temperature and the resistivity jump are
consistent with those in the literature \cite{PrCa,LB}.
Above the transition temperature, Mn$^{3+}$ and Mn$^{4+}$ are
randomly distributed, where the polaron energy is determined 
by the Jahn-Teller distortion of a single ion.
On the contrary, Mn$^{3+}$ and Mn$^{4+}$ are ordered below the 
transition temperature, where the polaron energy is modified
by the collective Jahn-Teller effect.

\begin{figure}[t]
\includegraphics[width=.8\linewidth]{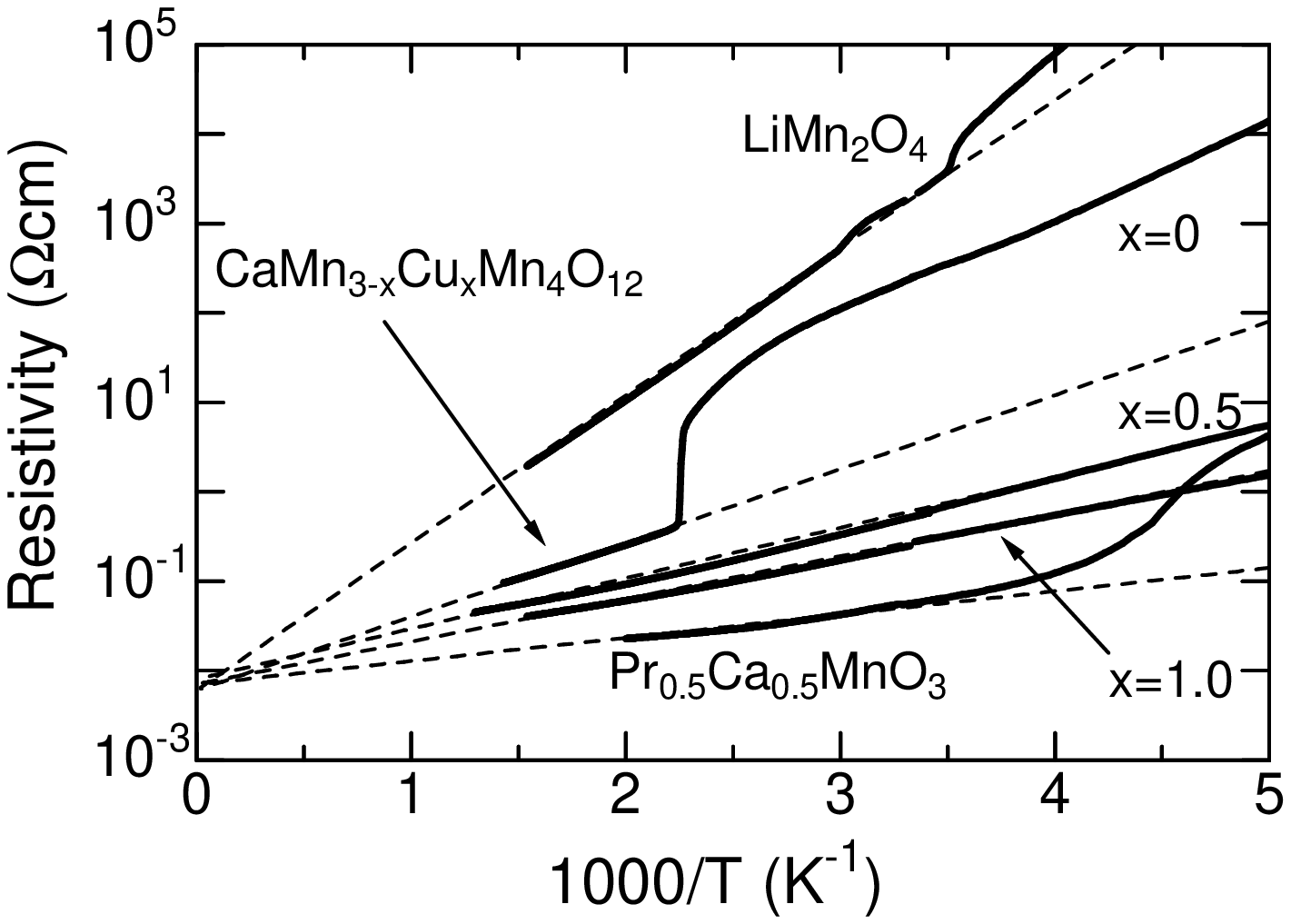}
\caption{
Temperature dependence of the resistivities
of various Mn oxides.
The dotted lines are fitting curves by Eq. (1).
}
\vspace*{5mm}
\includegraphics[width=.9\linewidth]{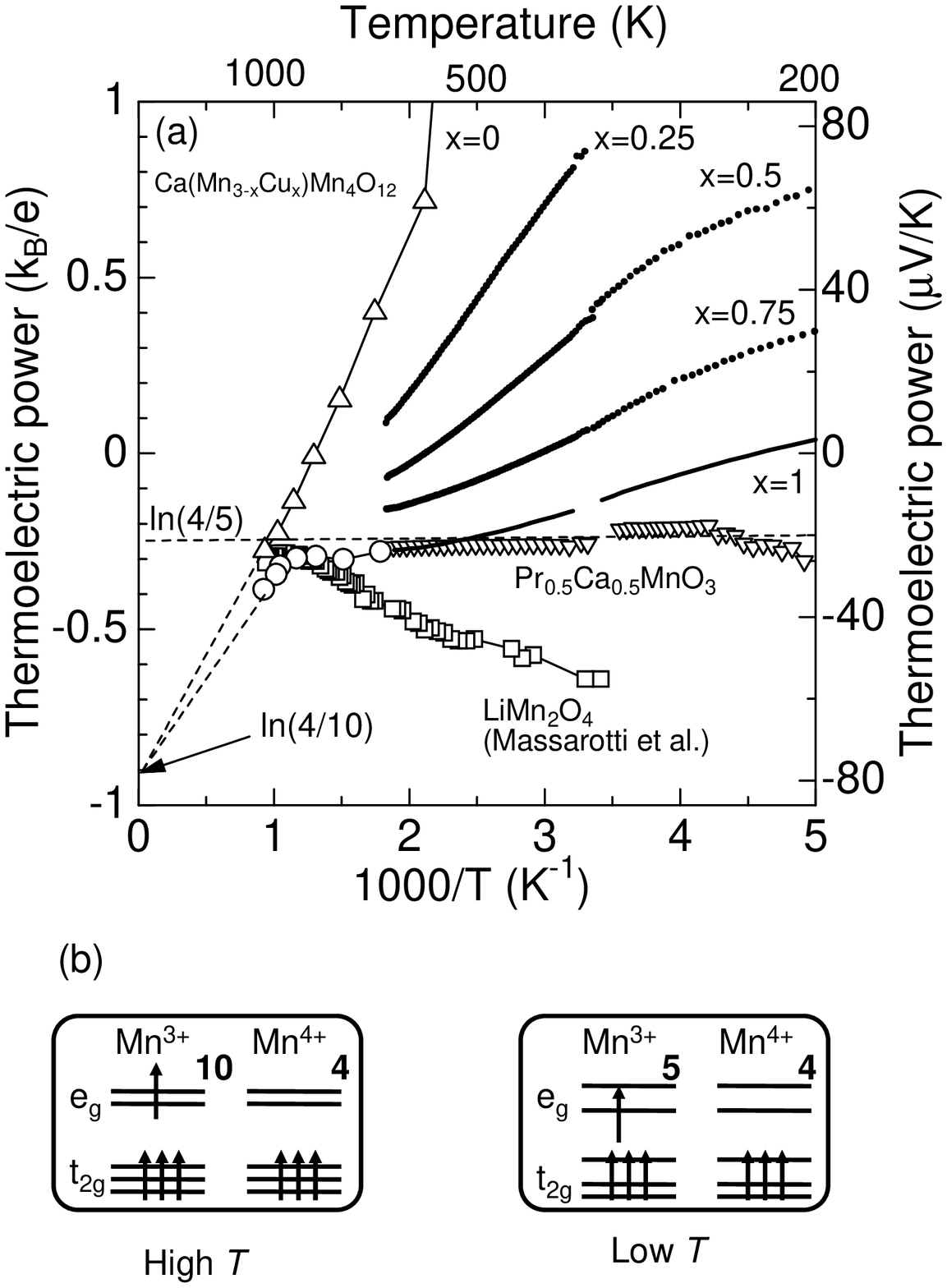}
\caption{
(a)Temperature dependence of the thermopowers
of various Mn oxides.
The dotted lines are fitting curves by Eq. (5).
The data for LiMn$_2$O$_4$ is taken from Ref. \cite{massa}.
(b) Electron configuration in Mn ions.
}
\end{figure}

We examined a least-square fitting with Eq. (1), as shown in
the dotted lines in Fig. 2.
Note that we used the data only above the transition temperature
for the charge-ordered samples.
As is clearly shown, the extrapolated values $\rho_0$ 
are in the range from 6 to 8 m$\Omega$cm.
Note that resistivity of solids can vary in a wide range
from 10$^{-6}$ (copper metal) to 10$^{20}$ (silica) $\Omega$cm.
In this sense, we can safely approximate 7$\pm$1 m$\Omega$cm 
as a constant value.

This result is quite anomalous.
Resistivity in the high temperature limit is known as the Ioffe-Regel limit 
(or Mott's minimum conductivity), where the mean free path is nearly equal 
to the lattice constant ($a$). The resistivity is equal to 
\begin{equation}
\rho=\frac{3\pi^2 \hbar}{e^2 ak_F^2},
\end{equation}
where $k_F$ is Fermi wave vector in a three-dimensional case \cite{mott}. 
Thus the high-temperature resistivity 
is dependent on carrier density 
and crystal structure through $k_F$ and $a$. 
Then $\rho_0$ is calculated to be 2 m$\Omega$cm
for $k_F$=$\pi/2a$ (Mn$^{3+}$:Mn$^{4+}$=1:1) 
and $a=$0.4 nm (a typical Mn-O-Mn distance of our samples), 
which is reasonably close to the observed value of 7$\pm$1 m$\Omega$cm.

Figure 3(a) shows the thermopowers of all the samples. 
All the data merge near $-25~\mu$V/K around 800-1000 K.
Note that the thermopower for LiMn$_2$O$_4$ is taken from Ref. \cite{massa}.
Koshibae {\textit et al.} \cite{koshibae} have proposed that
thermopower of the transition-metal oxides in the high temperature limit
is described by an extended Heikes formula written as
\begin{equation}
S=-\frac{k_B}{q}\ln \left(\frac{g_A}{g_{B}}\frac{p}{1-p}\right),
\end{equation}
where $g_i$ the degeneracy of the electron configuration on the $i$ ions, 
and $p$ the concentration of the B ion,
and $q$ is the charge difference between the A and B ions.
According to Eq. (3), we calculate the thermopower for 
Mn$^{3+}$:Mn$^{4+}$=1:1 ($p$=0.5) as
\begin{equation}
S=-\frac{k_B}{|e|}\ln \left(\frac{g_{{\rm Mn}^{4+}}}{g_{{\rm Mn}^{3+}}}\right).
\end{equation}
The Mn$^{3+}$ and Mn$^{4+}$ ions in the Mn oxides are in the high-spin state, 
where Mn$^{3+}$ has the electron configuration 
of ($t_{2g}$)$^3$($e_g$)$^1$ as shown in the inset of Fig. 2. 
Near 800 K, $E_{\rho}$ is larger than $k_BT$,
which implies that the Jahn-Teller distortion is effective for 
the polaron formation.
Thus the $e_g$ level splitting is valid  (no orbital degeneracy), 
and  $g_{{\rm Mn}^{3+}}$ should be equal to 5 
owing only to  the spin degeneracy of $S$=2 (see Fig. 3(b)).
Since the Mn$^{4+}$ ion [$(t_{2g})^3$] has no orbital degeneracy, 
$g_{{\rm Mn}^{4+}}$ is equal to 4 owing to $S$=3/2.
Then the thermopower is evaluated to be $-k_B\ln(5/4)/|e|=-20\mu$V/K,
which is in good agreement with the observed value of $-25~\mu$V/K.
Similar results for (La$_{1-x}$Ca$_x$)MnO$_3$ were reported 
by Palstra {\textit et al.}  \cite{palstra}. 

\begin{figure}[t]
\includegraphics[width=.8\linewidth]{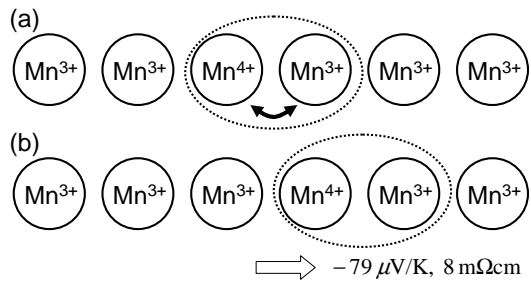}
\caption{
Schematic picture of electric conduction of Mn oxides
}
\end{figure}

Above 1000 K, the thermopowers begin to decrease, and seem to approach
$-79\pm 3~\mu$V/K as $T\to \infty$ \cite{kobaJPSJ}.
For high-temperature extrapolation, 
we assumed the expression for a high-temperature thermopower as
\begin{equation}
 S=S_0+\frac{k_B}{q}\frac{E_S}{k_BT},
\end{equation}
where $E_S$ ($E_{\rho}\gg E_S$) is a characteristic energy for $S$, 
and $S_0$ is the thermopower in the high-temperature limit
\cite{palstra}.
The physical meaning is simple:
the $e_g$-orbital degenaracy recovers above 1000 K, where
$g_{{\rm Mn}^{3+}}=10$ and $S=-k_B\ln(10/4)/|e|\sim -79~\mu$V/K
are expected, as shown in Fig. 3(b).

In the above argument, we evaluated the high-temperature 
resistivity and thermopower for Mn$^{3+}$:Mn$^{4+}$=1:1.
This is the case for  Ca(Mn$_2$Cu)Mn$_4$O$_{12}$, 
LiMn$_2$O$_4$, and Pr$_{0.5}$Ca$_{0.5}$MnO$_3$,
but is not applicable to Ca(Mn$_{3-x}$Cu$_x$)Mn$_4$O$_{12}$ ($x<$1).
However, the observed $\rho$ and $S$ seem to have universal values 
for $T\to \infty$ regardless of the Mn$^{4+}$ content or the carrier density.
We do not have a clear answer for this yet,
but propose a possible scenario 
that the electric conduction occurs in a highly localized way in the 
high temperature limit, where the exchange of entropy and charge occurs in 
the neighboring Mn$^{3+}$ and Mn$^{4+}$ ions. 
Figure 4 schematically shows a local exchange Mn$^{4+}$ for Mn$^{3+}$.
Suppose conduction occurs in a filamentary path of 
the Mn$^{4+}$ motion. 
Then the Mn$^{4+}$ ion induces the thermopower of $S=-k_B\ln(10/4)/|e|$
with $\rho_0=2$ m$\Omega$cm,
every time it replaces the Mn$^{3+}$ site. 
We think this phenomena would occur not only in the Mn oxides 
but also in other small-polaron conductors in general.

\section{Summary}
We prepared various Mn oxides 
consisting of Mn$^{3+}$ and Mn$^{4+}$ ions in the MnO$_6$ octahedron,
and measured their resistivity and thermopower from 4 to 1000 K.
Most unexpectedly, we have found that 
they seem to have unique asymptotic values in the high temperature limit. 
We have explained this in terms 
of the local ion exchange between Mn$^{3+}$ and Mn$^{4+}$. 

This work was partially supported by Grant-in-Aid for JSPS Fellows.

%\newpage
%\begin{center}
% Figure Captions
%\end{center}
%\begin{figure}[h]
%\caption{
%X-ray diffraction patters of
%(a) Ca(Mn$_2$Cu)Mn$_4$O$_{12}$ and (b) LiMn$_2$O$_4$.
%}

%\caption{
%Temperature dependence of the resistivities
%of various Mn oxides.
%The dotted lines are fitting curves by Eq. (1).
%}
%
%\caption{
%(a)Temperature dependence of the thermopowers
%of various Mn oxides.
%The dotted lines are fitting curves by Eq. (5).
%The data for LiMn$_2$O$_4$ is taken from Ref. \cite{massa}.
%(b) Electron configuration in Mn ions.
%}
%
%\caption{
%Schematic picture of electric conduction of Mn oxides
%}
%
%\end{figure}

\end{document}